\documentclass[aps,prl,twocolumn,superscriptaddress]{revtex4-1}  
\usepackage{graphicx}  
\graphicspath{{./Images_Def/}}
\usepackage{dcolumn}   
\usepackage{bm}        
\usepackage{amssymb, amsmath}   

\def\QE{{\sc Quantum ESPRESSO}}
\def\turboTDDFT{\texttt{turboTDDFT}}

\begin{document}


\title{Quantifying the plasmonic character of optical excitations in nanostructures} 

\author{Luca Bursi} 
\affiliation{Dipartimento di Fisica, Informatica e Matematica,
Universit\`a di Modena e Reggio Emilia, I-41125 Modena, Italy}
\affiliation{Istituto Nanoscienze CNR-NANO-S3, I-41125 Modena,
Italy}

\author{Arrigo Calzolari} \email{arrigo.calzolari@nano.cnr.it}
\affiliation{Istituto Nanoscienze CNR-NANO-S3, I-41125 Modena, Italy}

\author{Stefano Corni} \email{stefano.corni@nano.cnr.it}
\affiliation{Istituto Nanoscienze CNR-NANO-S3, I-41125 Modena, Italy}

\author{Elisa Molinari} 
\affiliation{Dipartimento di Fisica, Informatica e Matematica,
Universit\`a di Modena e Reggio Emilia, I-41125 Modena, Italy}
\affiliation{Istituto Nanoscienze CNR-NANO-S3, I-41125 Modena,
Italy}

\date{\today}

\begin{abstract}
The identification of plasmons in systems below $ \sim $10 nm in size is a tremendous challenge. Any sharp distinction of the excitation character (non-plasmonic vs plasmonic) becomes blurred in this range of sizes, where quantum effects become important. Here we define a {\em plasmonicicty index} that quantifies the plasmonic character of selected optical excitations in small nanostructures, starting from first principles calculations, based on (TD)DFT. This novel approach allows us to overcome the aforementioned problems, providing a direct and quantitative classification of the plasmonic character of the excitations. We show its usefulness for model metallic nanoparticles, a prototypical C-based molecule and a paradigmatic hybrid system. Our results indicate that the plasmonicity index can be exploited to solve previously unsolvable problems about the plasmonic character of complex systems, not predictable a priori.
\end{abstract}

\pacs{}
\maketitle

Localized surface plasmon resonances in nanostrucures interact strongly with light allowing the confinement of electromagnetic energy down to deep subwavelength regions\cite{Tame2013, Baumberg2012}. This, together with their easy tunability\cite{Morabito2015}, robustness\cite{GarciadeAbajo2015} and field enhancement properties\cite{AlonsoGonzalez2014}, provides a powerful tool to manipulate light at the nanoscale, below the diffraction limit. Thus, plasmons have become of paramount importance for a wide range of applications\cite{Halas2011, Li2015,Nordlander2014} spanning from light harvesting\cite{Atwater2010} to biosensing\cite{Mayer2011}.
In general terms, plasmons can be defined as electronic collective excitations that arise when the Coulomb interaction between excited states is switched on\cite{Bernadotte2013}. However, their theoretical description at the microscopic level is still an open and controversial issue\cite{FaradayDiscuss2015}.
In large nanoparticles optical and plasmonic properties are generally described by electrodynamics of continuous media, exploiting semiclassical models of the frequency-dependent dielectric function\cite{Wooten1972, KreibigVollmer, SalaDAgostino} and the identification of plasmons is straightforward. This description has been very useful for designing applications, but fails to convey a microscopic understanding of what plasmons are. Nanoparticles and their excitations are composed of electrons and nuclei like ordinary molecules. Therefore, it must be possible to understand their excited states, including plasmons, in terms of the same elementary electron and hole excitations routinely used to interpret molecular excited states. Notably, such a microscopic description is mandatory when the system size reaches 1-2 nanometers, where the dielectric description breaks down and quantum finite-size effects\cite{Baumberg2012, Thongrattanasiri2012} as well as the details of the atomic structure\cite{ZhangRubio2014} play a crucial role. However, at the nanoscale, single-particle and  plasmonic excitations are intrinsically mixed\cite{Nordlander2015}, and how to recognize a plasmonic excitation is still an unsolved problem.

A few approaches have been recently proposed attempting to classify the plasmonic character of the excitations of nanosystems.\cite{Bernadotte2013, Guidez2013, Townsend2014, Guidez2014Nanoscale, Bursi2014, Krauter2014, Townsend2015} In particular, Bernadotte et al.\cite{Bernadotte2013} formulated, in the framework of time-dependent density-functional theory (TDDFT), a scaling approach based on the different dependence of the energies of the excitations of nanosystems on the Coulomb kernel. 
Along this line, Krauter et al.\cite{Krauter2014} demonstrated that the electronic wave function of plasmons, at the time-dependent Hartree-Fock level, is described by the superposition of several electron configurations, i.e. Slater determinants, while this is not the case for non-plasmonic excitations. 

However, all the aforementioned approaches lack a simple quantification of the relative plasmonic character of the electronic excitations.
Frequently, especially at the nanoscale, a sharp classification of the excitations of a physical system in two categories, i.e. plasmonic or non-plasmonic, becomes ambiguous since plasmonic and non-plasmonic excitations with similar energies and similar symmetries mix\cite{Bernadotte2013}.
In this letter, we present an approach, based on the quantitative assessment of the plasmonic character of the excitations, which allows us to overcome this problem.
To this end, we define an index that quantifies the plasmonicity of a given excitation. In particular we will focus on light-induced optical excitations, that are the ones typically of interest in nanosystems.

When a monochromatic external scalar potential $ v_{ext}\left(\boldsymbol{r}, \omega\right) $ is applied to a physical system, its equilibrium charge density modifies as a response to the perturbation. This modification is the induced charge density 
$ n^\prime \left(\boldsymbol{r}, \omega\right)=
\int\chi \left(\boldsymbol{r},\boldsymbol{r}^\prime , \omega\right) v_{ext}\left(\boldsymbol{r}^\prime , \omega\right) d^3\boldsymbol{r}^\prime $ 
which in turn generates an induced potential 
$ v_{ind} \left(\boldsymbol{r}, \omega\right)=
\int f_{Coul}\left(\boldsymbol{r}-\boldsymbol{r}^\prime\right)n^\prime \left(\boldsymbol{r}^\prime , \omega\right) d^3\boldsymbol{r}^\prime $.
Here the external density response function $ \chi  $ and the Coulomb kernel $ f_{Coul}\left(\boldsymbol{r}-\boldsymbol{r}^\prime\right)=\frac{1}{\left|\boldsymbol{r}-\boldsymbol{r}^\prime\right|} $ have been introduced. The superposition of the external and the induced potential gives the total potential $ v_{tot}\left(\boldsymbol{r}, \omega\right)=v_{ext}\left(\boldsymbol{r}, \omega\right)+v_{ind}\left(\boldsymbol{r}, \omega\right) $ and $ n^\prime $ can be rewritten in terms of $ v_{tot} $ through the irreducible response function $ \chi _0 $ as 
$ n^\prime \left(\boldsymbol{r}, \omega\right)=
\int\chi _0\left(\boldsymbol{r},\boldsymbol{r}^\prime , \omega\right) v_{tot}\left(\boldsymbol{r}^\prime, \omega\right) d^3\boldsymbol{r}^\prime $. 
From the previous expressions, the well-known Dyson-like integral equation for the response functions can be obtained\cite{FetterWalecka} 
\begin{equation}
\begin{aligned}
\chi &\left(\boldsymbol{r},\boldsymbol{r}^\prime , \omega\right) =\chi _0\left(\boldsymbol{r},\boldsymbol{r}^\prime , \omega\right) +
\iint \chi _0\left(\boldsymbol{r},\boldsymbol{r}^{\prime\prime\prime} , \omega\right)  \\
& f_{Coul}\left(\boldsymbol{r}^{\prime\prime\prime}-\boldsymbol{r}^{\prime\prime}\right) \chi \left(\boldsymbol{r}^{\prime\prime},\boldsymbol{r}^\prime , \omega\right) d^3\boldsymbol{r}^{\prime\prime\prime} d^3\boldsymbol{r}^{\prime\prime} .
\end{aligned}
\label{eq:dyson}
\end{equation}
The poles $ \omega _\xi $  of the external response function, which correspond to the zero modes of $ \chi ^{-1} $,
\begin{equation}
\int \chi ^{-1}\left(\boldsymbol{r},\boldsymbol{r}^\prime , \omega _\xi\right)\rho _\xi\left(\boldsymbol{r}^\prime\right) d^3\boldsymbol{r}^\prime =0 ,
\label{eq:zero_modes}
\end{equation}
are the electronic excitations of the system. In Eq.\eqref{eq:zero_modes}, $ \rho $ is the transition density for the $\xi$ excitation.
Introducing the dielectric function $ \varepsilon $ \cite{MarquesGross, SalaDAgostino}, Eq.\eqref{eq:zero_modes} can be rewritten as 
$ \iint \varepsilon\left(\boldsymbol{r},\boldsymbol{r}^{\prime\prime}, \omega _\xi\right) \chi _0^{-1}\left(\boldsymbol{r}^{\prime\prime},\boldsymbol{r}^\prime , \omega _\xi\right) \rho _\xi\left(\boldsymbol{r}^\prime\right) d^3\boldsymbol{r}^{\prime\prime} d^3\boldsymbol{r}^\prime =0 $.
Considering the above equations and according to the typical classification adopted in solid-state physics\citep{Egri1985, FetterWalecka, Bernadotte2013}, the poles of $ \chi  $ can be divided into two types: (i) those corresponding to the zero modes of $ \varepsilon $, identified as plasmons and (ii) those originated from the poles of $ \chi _0 $, hereafter called non-plasmonic excitations.

However, the concept of $ \varepsilon $ for molecules and nanoparticles is not straightforward. To link the plasmonic character to a more intuitive quantity, here we follow an alternative approach. From Eq.\eqref{eq:dyson} we derive the relation
\begin{equation}
\chi ^{-1}\left(\boldsymbol{r},\boldsymbol{r}^\prime , \omega\right) = \chi _0^{-1}\left(\boldsymbol{r},\boldsymbol{r}^\prime , \omega\right) -f_{Coul}\left(\boldsymbol{r}-\boldsymbol{r}^\prime\right) .
\label{eq:chi_ext}
\end{equation}
Substituting Eq.\eqref{eq:chi_ext} in Eq.\eqref{eq:zero_modes}, we obtain 
\begin{equation}
\int \chi _0^{-1}\left(\boldsymbol{r},\boldsymbol{r}^\prime , \omega _\xi\right)\rho _\xi\left(\boldsymbol{r}^\prime\right) d^3\boldsymbol{r}^\prime - v_{ind \:\: \rho}\left(\boldsymbol{r}, \omega _\xi\right)=0 ,
\label{eq:main_eq}
\end{equation}
where $ v_{ind \:\: \rho} $ is the induced potential generated by $ \rho $.
In the same spirit of the analysis done above, we can classify the character of the excitation looking at the zeros of Eq.\eqref{eq:main_eq}. Non-plasmonic excitations correspond to the poles of the irreducible response function, i.e. to zero modes of $ \chi _0^{-1} $, and therefore at the frequency of an excitation of this kind, the first term of Eq.\eqref{eq:main_eq} vanishes. Since Eq.\eqref{eq:main_eq} still holds, also the induced potential in the case of a non-plasmonic excitation should in principles vanish. Plasmons, instead, do not correspond to zero modes of $ \chi _0^{-1} $, as we previously recalled, and thus the first term of Eq.\eqref{eq:main_eq} does not vanish, at the frequency of a plasmonic excitation. Therefore, the potential induced by a plasmon should remain finite, in order to satisfy Eq.\eqref{eq:main_eq}. The measure of how much the induced potential deviates from 0 can thus be interpreted as the measure of the plasmonic character of the excitation. We therefore introduce a {\em plasmonicity index} $ \eta _P $ as
\begin{equation}
\eta _P =
\displaystyle\frac{\int \left| v_{ind \:\: \rho}\left(\boldsymbol{r}, \omega _\xi\right) \right|^2d^3\boldsymbol{r}}
{\int \left| \rho _\xi\left(\boldsymbol{r}\right)\right|^2d^3\boldsymbol{r}}  = 
\frac{\int \left| \int \frac{\rho _\xi\left(\boldsymbol{r}^\prime \right)}{\left|\boldsymbol{r}-\boldsymbol{r}^\prime \right|} d^3\boldsymbol{r}^\prime\right|^2d^3\boldsymbol{r}}{\int \left| \rho _\xi\left(\boldsymbol{r} \right)\right|^2d^3\boldsymbol{r}} .
\label{eq:plasm_ind}
\end{equation}
In the light of the above considerations, the higher is $ \eta _P $, the more plasmonic is the excitation of frequency $ \omega _\xi $. The direct relation between the plasmonic nature of an excitation and the intensity of the relative induced potential is physically sound, as plasmons are typically connected with the local enhancement of the electromagnetic field. We choose the normalization of $ \eta _P $ in Eq.\eqref{eq:plasm_ind} to remove the trivial dependence of $ v_{ind \:\: \rho} $ on the normalization of $ \rho $. We also explored the use of an index with a different normalization\cite{SM}; the results, in line with those obtained by exploiting Eq.\eqref{eq:plasm_ind}, are given in Ref.\cite{SM}. 

Within this letter, we calculate the plasmonicity index for four paradigmatic molecular systems. We consider a linear Na$ _{20} $ chain and a tetrahedral Ag$ _{20} $ cluster as model plasmonic metallic nanosystems\cite{Bernadotte2013, Yan2007, Chen2013}, a naphthalene molecule as an example of molecular system that hosts ``molecular plasmons''\cite{Guidez2013, Manjavacas2013, Bursi2014, Lauchner2015} and a coupled system composed of a tetrahedral Ag$ _{20} $ cluster and a pyridine molecule which represents a prototypical hybrid system\cite{Zhao2006} mixing plasmonic and molecular portions. 
Their electronic structures and absorption spectra have been obtained from first principles simulations carried out by means of the \QE\cite{espresso} (QE) suite of codes, based on density-functional theory (DFT). We adopt the PBE\cite{PBE1996} \emph{Generalized Gradient Approximation} (GGA) to the \emph{exchange-correlation} (xc) functional and the electronic structure calculations are performed at the $ \Gamma $ point of the Brillouin zone. Within QE, wavefunctions and charge density are expanded in plane waves\cite{note_cutoffs} and the simulation exploits periodically repeated supercells, each containing the molecular system under study in central position and a suitable amount of vacuum ($12 $ \AA~ at least) to separate adjacent replica in the three spatial directions. This representation ensures also well converged optical spectra. The molecular structures of our systems are relaxed under the effect of the interatomic forces, with the exception of the hybrid system where the Ag$ _{20} $ cluster and pyridine were relaxed separately.
The optical absorption spectra and the response charge densities are computed exploiting the \turboTDDFT \! code\cite{Ge2014}, also part of the QE distribution, which implements, in the frequency domain, the Liouville-Lanczos approach to linearized TDDFT\cite{MarquesGross} and allows the calculation of the spectra in a relatively large energy range and in a computationally efficient way. This approach provides the induced densities $ n^\prime $ rather than the transition densities $ \rho _\xi $ required in Eq.\eqref{eq:plasm_ind}. We approximate the latter as a function of the former\cite{SM}.

Starting from the first principles results, we computed the plasmonicity index defined by Eq.\eqref{eq:plasm_ind}. 
When TDDFT methods are used, Eq.\eqref{eq:dyson} also contains the xc kernel, in addition to the Coulomb kernel. In this case it is still possible to relate plasmonic excitation to $ v_{ind \:\: \rho} $\cite{SM}.

\begin{figure}
\includegraphics[width=0.48\textwidth]{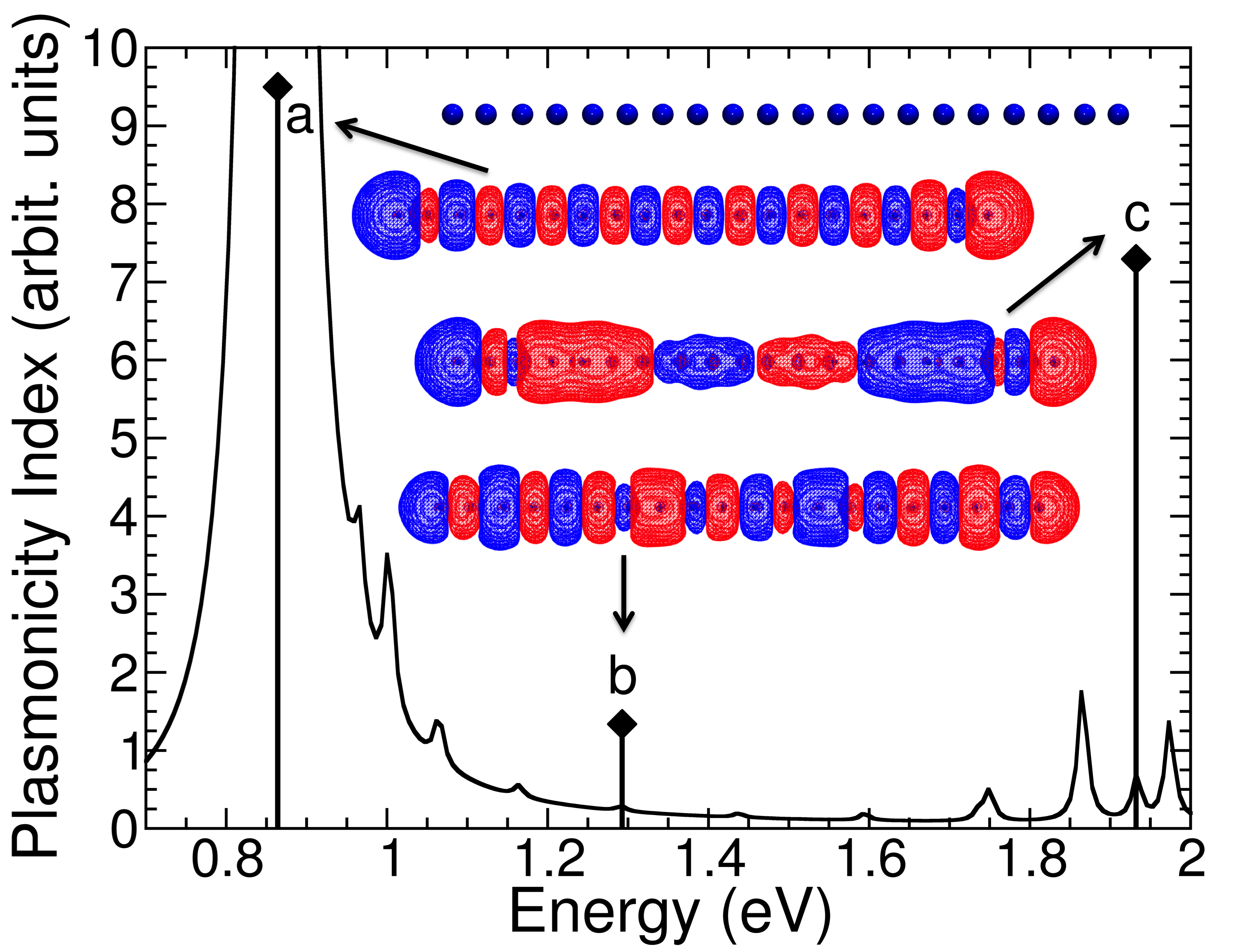}
\caption{\label{fig:Na_spectrum} (color online). TDDFT absorption spectrum (in arbitrary units) of Na$ _{20} $ chain (black line) and plasmonicity index (black diamonds and vertical lines) computed for selected peaks in the spectrum, labelled from a to c. The atomic structure of the chain and the imaginary parts of the TDDFT response charge densities computed at the frequencies of the peaks are shown as insets.}
\end{figure}

\emph{Na$ _{20} $}. In Fig.\ref{fig:Na_spectrum}, the TDDFT absorption spectrum (black solid line) of the Na$ _{20} $ chain (the atomic structure is shown as an inset) is dominated by an intense peak labelled with a in the low energy region. By using the same example as benchmark, Bernadotte et al.\cite{Bernadotte2013} have investigated the plasmonic properties of this system on the basis of the comparison with the 1D free-electron gas plasmons, identifying the transitions here labelled a and c as plasmons and peak b as non-plasmonic. We computed the TDDFT response charge density and the plasmonicity index for peaks a - c (Fig.\ref{fig:Na_spectrum}). Some more excitations are analysed in Ref.\cite{SM}. 
The response charge densities of a and b tend to be localized at the edges of the chain, leaving a charge depletion in the center, thus showing a dipolar character. This feature may indicate a plasmon resonance in a {\em quasi}-1D molecular system\cite{Bursi2014}, but it is still an ambiguous indication and it is not always sufficient to discriminate the plasmonic character. 
By plotting, instead, the values of the plasmonicity index (black diamonds in Fig.\ref{fig:Na_spectrum}) as a function of the energy, we obtain a ``plasmonic spectrum'' (black vertical lines in Fig.\ref{fig:Na_spectrum}) that spots plasmonic and non-plasmonic excitations. 
First of all, we note that the value of $ \eta _P $ is not trivially related to the oscillator strength of the peak. In particular, peak c results almost as plasmonic as peak a, despite its oscillator strength is negligible in comparison.
Secondly, the plasmonicity index of peaks a and c is markedly greater than that of peak b, identifying those excitations as plasmonic, in agreement with the independent assignment of Bernadotte et al.\cite{Bernadotte2013}. This is an important finding, as it shows that $ \eta _P $ correctly reproduces the results of more complex approaches, when they are applicable.

\begin{figure}
\includegraphics[width=0.48\textwidth]{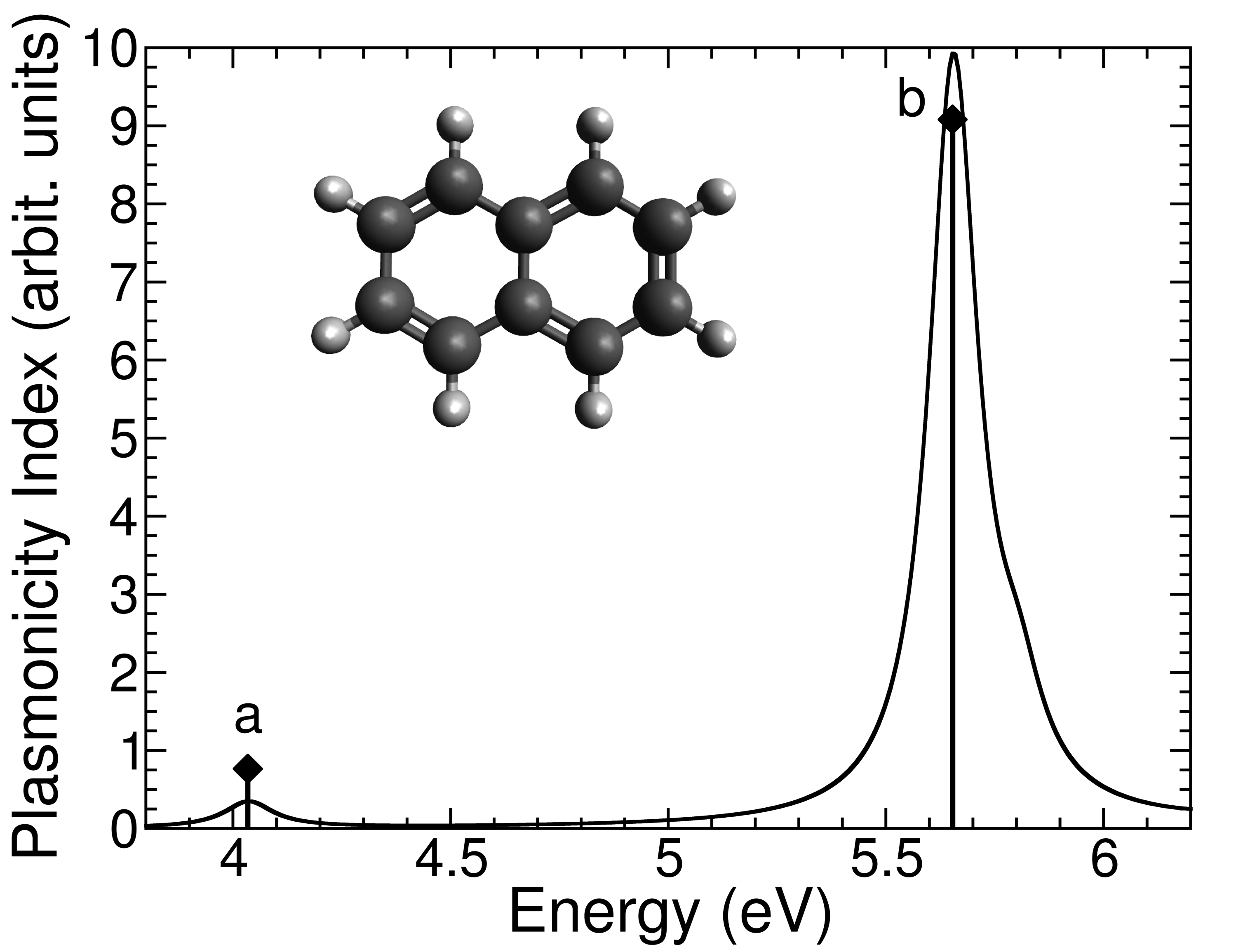}
\caption{\label{fig:Naft_spectrum} (color online). TDDFT absorption spectrum (in arbitrary units) of naphthalene (black line) in the low energy region and plasmonicity index (black diamonds and vertical lines) computed for peaks a and b in the spectrum. The molecular structure is shown as an inset.}
\end{figure}

\emph{Naphthalene}. Now we consider a carbon-based molecular system where only a few transitions can be observed in the low energy region of absorption. The spectrum of naphthalene in Fig.\ref{fig:Naft_spectrum} (where also the molecular structure is shown as inset) shows two bands, namely the most intense peak b and the weaker peak a at lower energy\cite{Guidez2013, Bursi2014}. We have studied the electronic, optical absorption, plasmonic and local field enhancement properties of this system elsewhere\cite{Bursi2014} with the same computational procedure and we refer to that for further details. In previous works, b has been identified as a molecular plasmon. The plasmonicity index analysis is in agreement with the picture just described, showing its usefulness also for truly non-metallic molecular systems.

\begin{figure}
\includegraphics[width=0.48\textwidth]{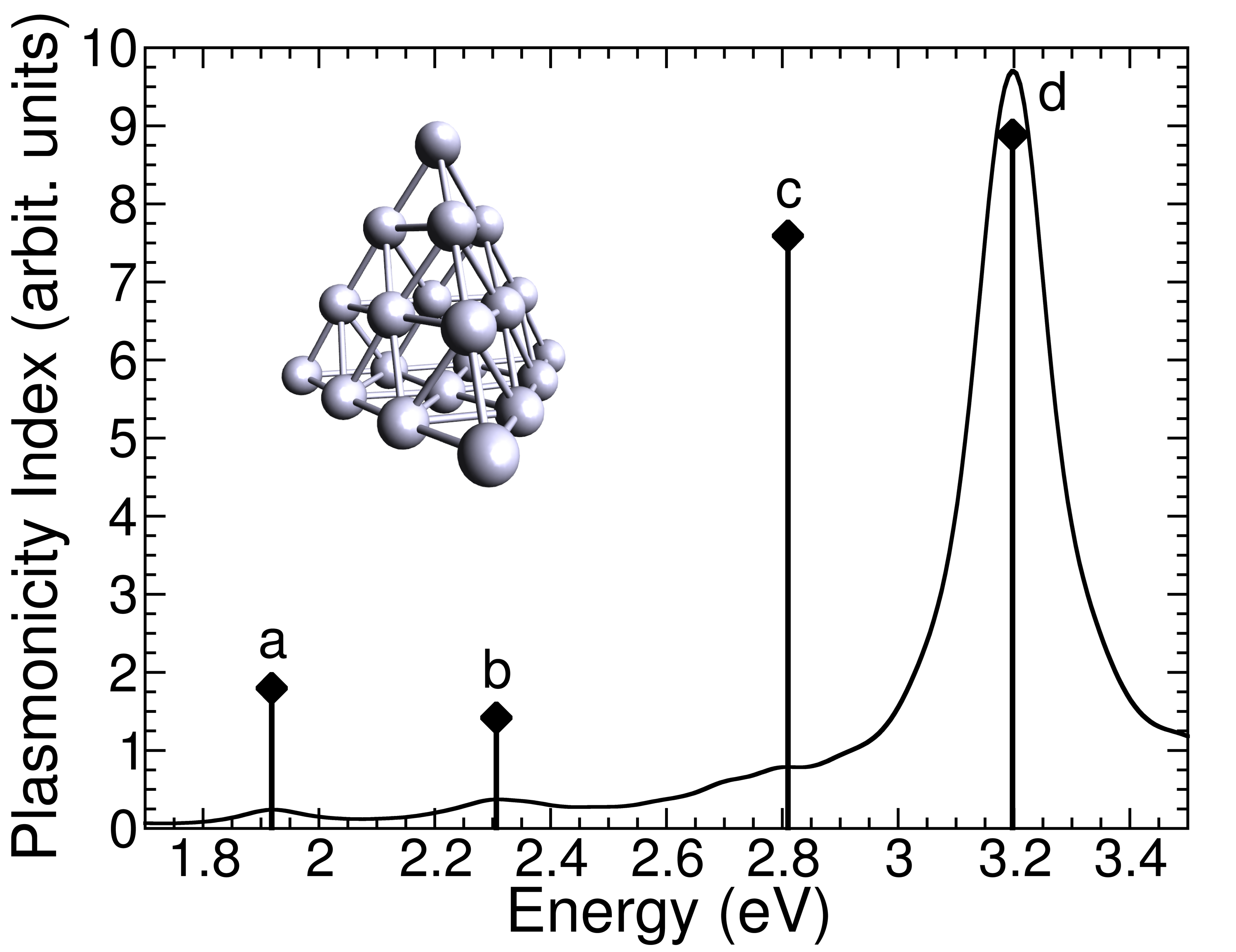}
\caption{\label{fig:Ag_spectrum} (color online). TDDFT absorption spectrum (in arbitrary units) of Ag$ _{20} $ cluster (black line) in the low energy region and plasmonicity index (black diamonds and vertical lines) computed for selected peaks in the spectrum, labelled from a to d. The atomic structure of the cluster is shown as an inset.}
\end{figure}

\emph{Ag$ _{20} $}. $ \eta _P $ is especially useful in the characterization of the plasmonic properties of systems where the visual inspection of the response charge density does not give straightforward indications, as the tetrahedral Ag$ _{20} $ cluster (the atomic structure is shown in the inset of Fig.\ref{fig:Ag_spectrum}). The latter is a well known plasmonic system\cite{Bernadotte2013, Chen2013, Zhao2006}, despite its small number of atoms. An intense peak dominates the low energy region of the adsorption spectrum of this cluster (see Fig.\ref{fig:Ag_spectrum}) and, according to Bernadotte et al.\cite{Bernadotte2013}, in correspondence to that peak, plasmonic and non-plasmonic excitations mix not allowing an easy identification of their character. In this case, the most plasmonic excitation, according to the index, is the one responsible for the most intense peak d, as expected, but also peak c shows an intense plasmonic character. This can be interpreted as a consequence of the mixed character of the excitation c, as discussed by Bernadotte et al.\cite{Bernadotte2013}. But most importantly, this result demonstrates that the plasmonicity index provides a quantitative classification of the excitations which works also for an excitation showing a mixed character.

\begin{figure}
\includegraphics[width=0.48\textwidth]{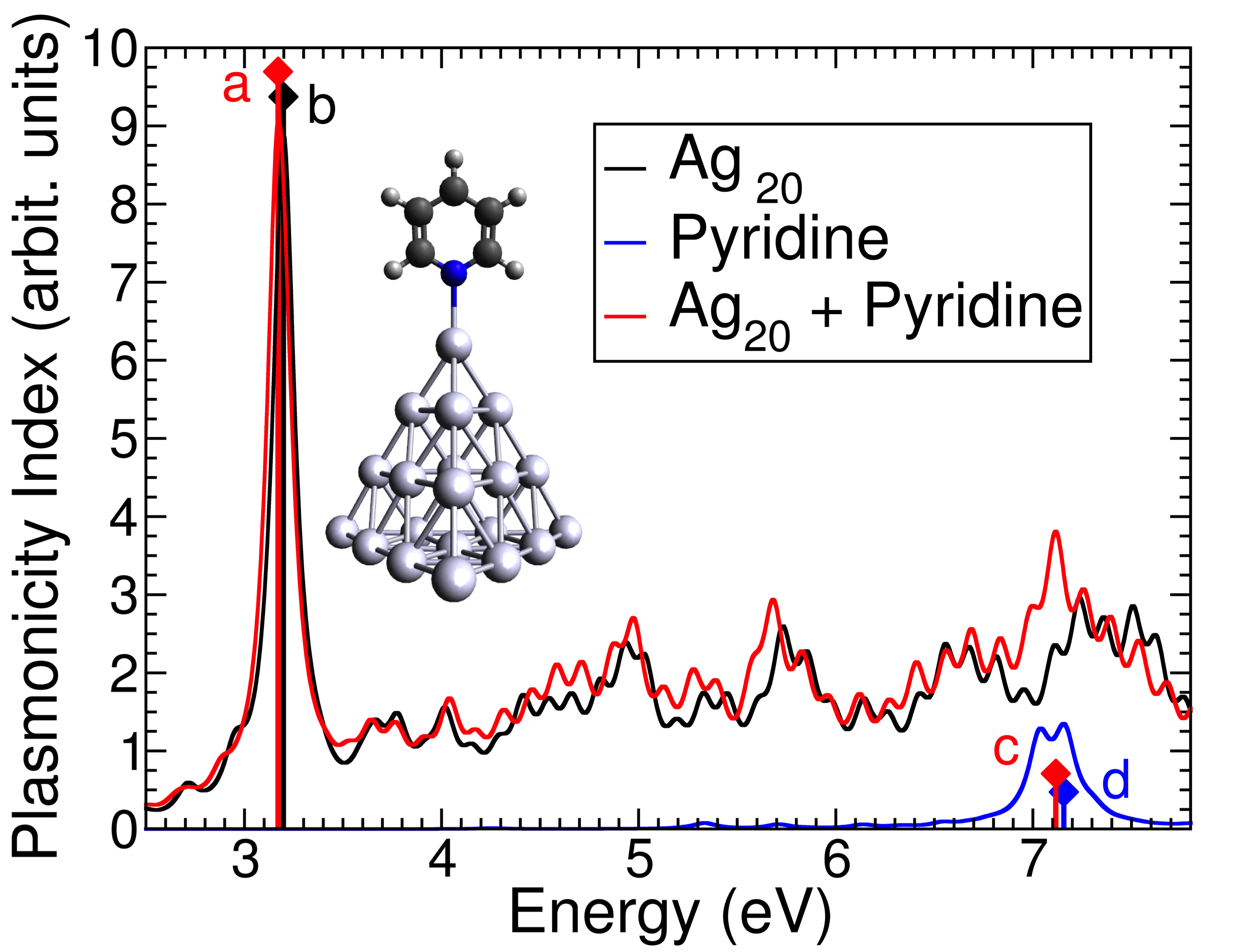}
\caption{\label{fig:Ag_Py_spectrum} (color online). TDDFT absorption spectra (in arbitrary units) of Ag$ _{20} $ cluster (black line), pyridine molecule (blue line) and the coupled hybrid system Ag$ _{20} $  $ + $ pyridine (red line). The plasmonicity index (diamonds and vertical lines), computed for the main peaks in the spectra of these three systems are depicted and labelled from a to d with the same color scheme for clarity. The atomic structure of Ag$ _{20} $  $ + $ pyridine is shown as an inset.}
\end{figure}

\emph{Ag$ _{20} $  $ + $ pyridine}. One of the objectives of the definition of $ \eta _P $ is its application to hybrid systems mixing plasmonic and molecular components, where the analysis of the plasmonic properties is often not straightforward. To this aim, we considered Ag$ _{20} $ coupled to a pyridine molecule (the atomic structure is shown in the inset of Fig.\ref{fig:Ag_Py_spectrum}), which is a venerable model system for surface enhanced spectroscopy\cite{Zhao2006}. 
The TDDFT absorption spectra of Ag$ _{20} $, pyridine and Ag$ _{20} $  $ + $ pyridine are shown in Fig.\ref{fig:Ag_Py_spectrum} together with the plasmonicity index computed only for the most intense peaks in the spectra. 

The index of the main absorption peak of the Ag$ _{20} $ cluster alone (peak b) is much larger compared to the index of the main absorption peak of pyridine alone (peak d). Thus, the classification in terms of the plasmonic character of the excitations provided by $ \eta _P $ for the isolated systems is the intuitive one.
When we compute the index for the corresponding peaks in the hybrid system (peaks a and c, respectively), we find that the intuitive classification is maintained, namely $ \eta _P $ of a is close to $ \eta _P $ of b and $ \eta _P $ of c is close to $ \eta _P $ of d, as expected.

In conclusion, we defined an index which quantifies the plasmonic character of the excitations in nanostructures exploiting directly the results of first principles simulations. Within this letter, we validated this plasmonicity index $ \eta _P $ on a Na$ _{20} $ chain and on a naphthalene. We then used the index to characterize plasmonic properties of a Ag$ _{20} $ cluster, where excitations of a mixed character have been observed, and for a paradigmatic hybrid system. The results provided by $ \eta _P $ allow to gain insights into the microscopic origin of the plasmonic resonances in small isolated and hybrid nanostructures, thus paving the way for applications to more complex systems whose plasmonic properties are not easily predictable a priori.

Computer resources were provided by the CINECA supercomputing center at their Fermi BG/Q machine through the \emph{ISCRA C} \emph{PlasmInd} project.
The authors would like to thank in particular Nicola Spallanzani at CINECA for computational support. 

\bibliographystyle{apsrev4-1}
\bibliography{bursi_bib}

\begin{thebibliography}{37}%
\makeatletter
\providecommand \@ifxundefined [1]{%
 \@ifx{#1\undefined}
}%
\providecommand \@ifnum [1]{%
 \ifnum #1\expandafter \@firstoftwo
 \else \expandafter \@secondoftwo
 \fi
}%
\providecommand \@ifx [1]{%
 \ifx #1\expandafter \@firstoftwo
 \else \expandafter \@secondoftwo
 \fi
}%
\providecommand \natexlab [1]{#1}%
\providecommand \enquote  [1]{``#1''}%
\providecommand \bibnamefont  [1]{#1}%
\providecommand \bibfnamefont [1]{#1}%
\providecommand \citenamefont [1]{#1}%
\providecommand \href@noop [0]{\@secondoftwo}%
\providecommand \href [0]{\begingroup \@sanitize@url \@href}%
\providecommand \@href[1]{\@@startlink{#1}\@@href}%
\providecommand \@@href[1]{\endgroup#1\@@endlink}%
\providecommand \@sanitize@url [0]{\catcode `\\12\catcode `\$12\catcode
  `\&12\catcode `\#12\catcode `\^12\catcode `\_12\catcode `\%12\relax}%
\providecommand \@@startlink[1]{}%
\providecommand \@@endlink[0]{}%
\providecommand \url  [0]{\begingroup\@sanitize@url \@url }%
\providecommand \@url [1]{\endgroup\@href {#1}{\urlprefix }}%
\providecommand \urlprefix  [0]{URL }%
\providecommand \Eprint [0]{\href }%
\providecommand \doibase [0]{http://dx.doi.org/}%
\providecommand \selectlanguage [0]{\@gobble}%
\providecommand \bibinfo  [0]{\@secondoftwo}%
\providecommand \bibfield  [0]{\@secondoftwo}%
\providecommand \translation [1]{[#1]}%
\providecommand \BibitemOpen [0]{}%
\providecommand \bibitemStop [0]{}%
\providecommand \bibitemNoStop [0]{.\EOS\space}%
\providecommand \EOS [0]{\spacefactor3000\relax}%
\providecommand \BibitemShut  [1]{\csname bibitem#1\endcsname}%
\let\auto@bib@innerbib\@empty
\bibitem [{\citenamefont {Tame}\ \emph {et~al.}(2013)\citenamefont {Tame},
  \citenamefont {McEnery}, \citenamefont {\"Ozdemir}, \citenamefont {Lee},
  \citenamefont {Maier},\ and\ \citenamefont {Kim}}]{Tame2013}%
  \BibitemOpen
  \bibfield  {author} {\bibinfo {author} {\bibfnamefont {M.~S.}\ \bibnamefont
  {Tame}}, \bibinfo {author} {\bibfnamefont {K.~R.}\ \bibnamefont {McEnery}},
  \bibinfo {author} {\bibfnamefont {{\c{S}}.~K.}\ \bibnamefont {\"Ozdemir}},
  \bibinfo {author} {\bibfnamefont {J.}~\bibnamefont {Lee}}, \bibinfo {author}
  {\bibfnamefont {S.~A.}\ \bibnamefont {Maier}}, \ and\ \bibinfo {author}
  {\bibfnamefont {M.~S.}\ \bibnamefont {Kim}},\ }\href@noop {} {\bibfield
  {journal} {\bibinfo  {journal} {Nat. Phys.}\ }\textbf {\bibinfo {volume}
  {9}},\ \bibinfo {pages} {329} (\bibinfo {year} {2013})}\BibitemShut {NoStop}%
\bibitem [{\citenamefont {Savage}\ \emph {et~al.}(2012)\citenamefont {Savage},
  \citenamefont {Hawkeye}, \citenamefont {Esteban}, \citenamefont {Borisov},
  \citenamefont {Aizpurua},\ and\ \citenamefont {Baumberg}}]{Baumberg2012}%
  \BibitemOpen
  \bibfield  {author} {\bibinfo {author} {\bibfnamefont {K.~J.}\ \bibnamefont
  {Savage}}, \bibinfo {author} {\bibfnamefont {M.~M.}\ \bibnamefont {Hawkeye}},
  \bibinfo {author} {\bibfnamefont {R.}~\bibnamefont {Esteban}}, \bibinfo
  {author} {\bibfnamefont {A.~G.}\ \bibnamefont {Borisov}}, \bibinfo {author}
  {\bibfnamefont {J.}~\bibnamefont {Aizpurua}}, \ and\ \bibinfo {author}
  {\bibfnamefont {J.~J.}\ \bibnamefont {Baumberg}},\ }\href@noop {} {\bibfield
  {journal} {\bibinfo  {journal} {Nature}\ }\textbf {\bibinfo {volume} {491}},\
  \bibinfo {pages} {574Ð577} (\bibinfo {year} {2012})}\BibitemShut {NoStop}%
\bibitem [{\citenamefont {Linic}\ \emph {et~al.}(2015)\citenamefont {Linic},
  \citenamefont {andCalvin Boerigter},\ and\ \citenamefont
  {Morabito}}]{Morabito2015}%
  \BibitemOpen
  \bibfield  {author} {\bibinfo {author} {\bibfnamefont {S.}~\bibnamefont
  {Linic}}, \bibinfo {author} {\bibfnamefont {U.~A.}\ \bibnamefont {andCalvin
  Boerigter}}, \ and\ \bibinfo {author} {\bibfnamefont {M.}~\bibnamefont
  {Morabito}},\ }\href@noop {} {\bibfield  {journal} {\bibinfo  {journal}
  {Nature Mater.}\ }\textbf {\bibinfo {volume} {14}},\ \bibinfo {pages}
  {567Ð576} (\bibinfo {year} {2015})}\BibitemShut {NoStop}%
\bibitem [{\citenamefont {{Garc\'{\i}a de Abajo}}\ and\ \citenamefont
  {Manjavacas}(2015)}]{GarciadeAbajo2015}%
  \BibitemOpen
  \bibfield  {author} {\bibinfo {author} {\bibfnamefont {F.~J.}\ \bibnamefont
  {{Garc\'{\i}a de Abajo}}}\ and\ \bibinfo {author} {\bibfnamefont
  {A.}~\bibnamefont {Manjavacas}},\ }\href {\doibase 10.1039/C4FD00216D}
  {\bibfield  {journal} {\bibinfo  {journal} {Faraday Discuss.}\ }\textbf
  {\bibinfo {volume} {178}},\ \bibinfo {pages} {87 } (\bibinfo {year}
  {2015})}\BibitemShut {NoStop}%
\bibitem [{\citenamefont {Alonso-Gonz\'{a}lez}\ \emph
  {et~al.}(2014)\citenamefont {Alonso-Gonz\'{a}lez}, \citenamefont {Nikitin},
  \citenamefont {Golmar}, \citenamefont {Centeno}, \citenamefont {Pesquera},
  \citenamefont {V\'{e}lez}, \citenamefont {Chen}, \citenamefont {Navickaite},
  \citenamefont {Koppens}, \citenamefont {Zurutuza}, \citenamefont {Casanova},
  \citenamefont {Hueso},\ and\ \citenamefont
  {Hillenbrand}}]{AlonsoGonzalez2014}%
  \BibitemOpen
  \bibfield  {author} {\bibinfo {author} {\bibfnamefont {P.}~\bibnamefont
  {Alonso-Gonz\'{a}lez}}, \bibinfo {author} {\bibfnamefont {A.~Y.}\
  \bibnamefont {Nikitin}}, \bibinfo {author} {\bibfnamefont {F.}~\bibnamefont
  {Golmar}}, \bibinfo {author} {\bibfnamefont {A.}~\bibnamefont {Centeno}},
  \bibinfo {author} {\bibfnamefont {A.}~\bibnamefont {Pesquera}}, \bibinfo
  {author} {\bibfnamefont {S.}~\bibnamefont {V\'{e}lez}}, \bibinfo {author}
  {\bibfnamefont {J.}~\bibnamefont {Chen}}, \bibinfo {author} {\bibfnamefont
  {G.}~\bibnamefont {Navickaite}}, \bibinfo {author} {\bibfnamefont
  {F.}~\bibnamefont {Koppens}}, \bibinfo {author} {\bibfnamefont
  {A.}~\bibnamefont {Zurutuza}}, \bibinfo {author} {\bibfnamefont
  {F.}~\bibnamefont {Casanova}}, \bibinfo {author} {\bibfnamefont {L.~E.}\
  \bibnamefont {Hueso}}, \ and\ \bibinfo {author} {\bibfnamefont
  {R.}~\bibnamefont {Hillenbrand}},\ }\href
  {http://www.scopus.com/inward/record.url?eid=2-s2.0-84903579704&partnerID=40&md5=9dc79b507dfcaf2e22d1ed4357276ad5}
  {\bibfield  {journal} {\bibinfo  {journal} {Science}\ }\textbf {\bibinfo
  {volume} {344}},\ \bibinfo {pages} {1369} (\bibinfo {year}
  {2014})}\BibitemShut {NoStop}%
\bibitem [{\citenamefont {Halas}\ \emph {et~al.}(2011)\citenamefont {Halas},
  \citenamefont {Lal}, \citenamefont {Chang}, \citenamefont {Link},\ and\
  \citenamefont {Nordlander}}]{Halas2011}%
  \BibitemOpen
  \bibfield  {author} {\bibinfo {author} {\bibfnamefont {N.~J.}\ \bibnamefont
  {Halas}}, \bibinfo {author} {\bibfnamefont {S.}~\bibnamefont {Lal}}, \bibinfo
  {author} {\bibfnamefont {W.-S.}\ \bibnamefont {Chang}}, \bibinfo {author}
  {\bibfnamefont {S.}~\bibnamefont {Link}}, \ and\ \bibinfo {author}
  {\bibfnamefont {P.}~\bibnamefont {Nordlander}},\ }\href {\doibase
  10.1021/cr200061k} {\bibfield  {journal} {\bibinfo  {journal} {Chem. Rev.}\
  }\textbf {\bibinfo {volume} {111}},\ \bibinfo {pages} {3913} (\bibinfo {year}
  {2011})}\BibitemShut {NoStop}%
\bibitem [{\citenamefont {Li}\ \emph {et~al.}(2015)\citenamefont {Li},
  \citenamefont {Coppens}, \citenamefont {Besteiro}, \citenamefont {Wang},
  \citenamefont {Govorov},\ and\ \citenamefont {Valentine}}]{Li2015}%
  \BibitemOpen
  \bibfield  {author} {\bibinfo {author} {\bibfnamefont {W.}~\bibnamefont
  {Li}}, \bibinfo {author} {\bibfnamefont {Z.~J.}\ \bibnamefont {Coppens}},
  \bibinfo {author} {\bibfnamefont {L.~V.}\ \bibnamefont {Besteiro}}, \bibinfo
  {author} {\bibfnamefont {W.}~\bibnamefont {Wang}}, \bibinfo {author}
  {\bibfnamefont {A.~O.}\ \bibnamefont {Govorov}}, \ and\ \bibinfo {author}
  {\bibfnamefont {J.}~\bibnamefont {Valentine}},\ }\href@noop {} {\bibfield
  {journal} {\bibinfo  {journal} {Nature Comm.}\ }\textbf {\bibinfo {volume}
  {6}},\ \bibinfo {pages} {8379} (\bibinfo {year} {2015})}\BibitemShut
  {NoStop}%
\bibitem [{\citenamefont {Zheng}\ \emph {et~al.}(2014)\citenamefont {Zheng},
  \citenamefont {Wang}, \citenamefont {Nordlander},\ and\ \citenamefont
  {Halas}}]{Nordlander2014}%
  \BibitemOpen
  \bibfield  {author} {\bibinfo {author} {\bibfnamefont {B.~Y.}\ \bibnamefont
  {Zheng}}, \bibinfo {author} {\bibfnamefont {Y.}~\bibnamefont {Wang}},
  \bibinfo {author} {\bibfnamefont {P.}~\bibnamefont {Nordlander}}, \ and\
  \bibinfo {author} {\bibfnamefont {N.~J.}\ \bibnamefont {Halas}},\ }\href@noop
  {} {\bibfield  {journal} {\bibinfo  {journal} {Adv. Mater.}\ }\textbf
  {\bibinfo {volume} {26}},\ \bibinfo {pages} {6318Ð6323} (\bibinfo {year}
  {2014})}\BibitemShut {NoStop}%
\bibitem [{\citenamefont {Atwater}\ and\ \citenamefont
  {Polman}(2010)}]{Atwater2010}%
  \BibitemOpen
  \bibfield  {author} {\bibinfo {author} {\bibfnamefont {H.~A.}\ \bibnamefont
  {Atwater}}\ and\ \bibinfo {author} {\bibfnamefont {A.}~\bibnamefont
  {Polman}},\ }\href
  {http://www.scopus.com/inward/record.url?eid=2-s2.0-77249099338&partnerID=40&md5=9abe7940d9b299804510d4299b35206f}
  {\bibfield  {journal} {\bibinfo  {journal} {Nat. Mater.}\ }\textbf {\bibinfo
  {volume} {9}},\ \bibinfo {pages} {205} (\bibinfo {year} {2010})}\BibitemShut
  {NoStop}%
\bibitem [{\citenamefont {Mayer}\ and\ \citenamefont
  {Hafner}(2011)}]{Mayer2011}%
  \BibitemOpen
  \bibfield  {author} {\bibinfo {author} {\bibfnamefont {K.~M.}\ \bibnamefont
  {Mayer}}\ and\ \bibinfo {author} {\bibfnamefont {J.~H.}\ \bibnamefont
  {Hafner}},\ }\href {\doibase 10.1021/cr100313v} {\bibfield  {journal}
  {\bibinfo  {journal} {Chem. Rev.}\ }\textbf {\bibinfo {volume} {111}},\
  \bibinfo {pages} {3828} (\bibinfo {year} {2011})}\BibitemShut {NoStop}%
\bibitem [{\citenamefont {Bernadotte}\ \emph {et~al.}(2013)\citenamefont
  {Bernadotte}, \citenamefont {Evers},\ and\ \citenamefont
  {Jacob}}]{Bernadotte2013}%
  \BibitemOpen
  \bibfield  {author} {\bibinfo {author} {\bibfnamefont {S.}~\bibnamefont
  {Bernadotte}}, \bibinfo {author} {\bibfnamefont {F.}~\bibnamefont {Evers}}, \
  and\ \bibinfo {author} {\bibfnamefont {C.~R.}\ \bibnamefont {Jacob}},\ }\href
  {\doibase 10.1021/jp3113073} {\bibfield  {journal} {\bibinfo  {journal} {J.
  Phys. Chem. C}\ }\textbf {\bibinfo {volume} {117}},\ \bibinfo {pages} {1863}
  (\bibinfo {year} {2013})}\BibitemShut {NoStop}%
\bibitem [{\citenamefont {de~Abajo}\ \emph {et~al.}(2015)\citenamefont
  {de~Abajo}, \citenamefont {Sapienza}, \citenamefont {Noginov}, \citenamefont
  {Benz}, \citenamefont {Baumberg}, \citenamefont {Maier}, \citenamefont
  {Graham}, \citenamefont {Aizpurua}, \citenamefont {Ebbesen}, \citenamefont
  {Pinchuk}, \citenamefont {Khurgin}, \citenamefont {Matczyszyn}, \citenamefont
  {Hugall}, \citenamefont {van Hulst}, \citenamefont {Dawson}, \citenamefont
  {Roberts}, \citenamefont {Nielsen}, \citenamefont {Bursi}, \citenamefont
  {Flatt\'{e}}, \citenamefont {Yi}, \citenamefont {Hess}, \citenamefont
  {Engheta}, \citenamefont {Brongersma}, \citenamefont {Podolskiy},
  \citenamefont {Shalaev}, \citenamefont {Narimanov},\ and\ \citenamefont
  {Zayats}}]{FaradayDiscuss2015}%
  \BibitemOpen
  \bibfield  {author} {\bibinfo {author} {\bibfnamefont {F.~J.~G.}\
  \bibnamefont {de~Abajo}}, \bibinfo {author} {\bibfnamefont {R.}~\bibnamefont
  {Sapienza}}, \bibinfo {author} {\bibfnamefont {M.}~\bibnamefont {Noginov}},
  \bibinfo {author} {\bibfnamefont {F.}~\bibnamefont {Benz}}, \bibinfo {author}
  {\bibfnamefont {J.}~\bibnamefont {Baumberg}}, \bibinfo {author}
  {\bibfnamefont {S.}~\bibnamefont {Maier}}, \bibinfo {author} {\bibfnamefont
  {D.}~\bibnamefont {Graham}}, \bibinfo {author} {\bibfnamefont
  {J.}~\bibnamefont {Aizpurua}}, \bibinfo {author} {\bibfnamefont
  {T.}~\bibnamefont {Ebbesen}}, \bibinfo {author} {\bibfnamefont
  {A.}~\bibnamefont {Pinchuk}}, \bibinfo {author} {\bibfnamefont
  {J.}~\bibnamefont {Khurgin}}, \bibinfo {author} {\bibfnamefont
  {K.}~\bibnamefont {Matczyszyn}}, \bibinfo {author} {\bibfnamefont {J.~T.}\
  \bibnamefont {Hugall}}, \bibinfo {author} {\bibfnamefont {N.}~\bibnamefont
  {van Hulst}}, \bibinfo {author} {\bibfnamefont {P.}~\bibnamefont {Dawson}},
  \bibinfo {author} {\bibfnamefont {C.}~\bibnamefont {Roberts}}, \bibinfo
  {author} {\bibfnamefont {M.}~\bibnamefont {Nielsen}}, \bibinfo {author}
  {\bibfnamefont {L.}~\bibnamefont {Bursi}}, \bibinfo {author} {\bibfnamefont
  {M.}~\bibnamefont {Flatt\'{e}}}, \bibinfo {author} {\bibfnamefont
  {J.}~\bibnamefont {Yi}}, \bibinfo {author} {\bibfnamefont {O.}~\bibnamefont
  {Hess}}, \bibinfo {author} {\bibfnamefont {N.}~\bibnamefont {Engheta}},
  \bibinfo {author} {\bibfnamefont {M.}~\bibnamefont {Brongersma}}, \bibinfo
  {author} {\bibfnamefont {V.}~\bibnamefont {Podolskiy}}, \bibinfo {author}
  {\bibfnamefont {V.}~\bibnamefont {Shalaev}}, \bibinfo {author} {\bibfnamefont
  {E.}~\bibnamefont {Narimanov}}, \ and\ \bibinfo {author} {\bibfnamefont
  {A.}~\bibnamefont {Zayats}},\ }\href {\doibase 10.1039/c5fd90022k} {\bibfield
   {journal} {\bibinfo  {journal} {Faraday Discuss.}\ }\textbf {\bibinfo
  {volume} {178}},\ \bibinfo {pages} {123} (\bibinfo {year}
  {2015})}\BibitemShut {NoStop}%
\bibitem [{Woo()}]{Wooten1972}%
  \BibitemOpen
  \href@noop {} {}\bibinfo {note} {F. Wooten, {\em Optical Properties of
  solids} (Academic Press, Inc. San Diego CA, 9211 USA, 1972).}\BibitemShut
  {Stop}%
\bibitem [{Kre()}]{KreibigVollmer}%
  \BibitemOpen
  \href@noop {} {}\bibinfo {note} {U. Kreibig and M. Vollmer, {\em Optical
  Properties of Metal Clusters}, Springer series in material science, Vol. 25
  (Springer, Berlin, 1995).}\BibitemShut {Stop}%
\bibitem [{Sal()}]{SalaDAgostino}%
  \BibitemOpen
  \href@noop {} {}\bibinfo {note} {F. {della Sala} and S. D'Agostino, {\em
  Handbook of Molecular Plasmonics} (Pan Stanford Publishing Pte. Ltd.,
  2013)}\BibitemShut {NoStop}%
\bibitem [{\citenamefont {Thongrattanasiri}\ \emph {et~al.}(2012)\citenamefont
  {Thongrattanasiri}, \citenamefont {Manjavacas},\ and\ \citenamefont
  {Garc\'{i}a~de Abajo}}]{Thongrattanasiri2012}%
  \BibitemOpen
  \bibfield  {author} {\bibinfo {author} {\bibfnamefont {S.}~\bibnamefont
  {Thongrattanasiri}}, \bibinfo {author} {\bibfnamefont {A.}~\bibnamefont
  {Manjavacas}}, \ and\ \bibinfo {author} {\bibfnamefont {F.~J.}\ \bibnamefont
  {Garc\'{i}a~de Abajo}},\ }\href
  {http://www.scopus.com/inward/record.url?eid=2-s2.0-84857778581&partnerID=40&md5=434d441be7254d3e11774d498998aa45}
  {\bibfield  {journal} {\bibinfo  {journal} {ACS Nano}\ }\textbf {\bibinfo
  {volume} {6}},\ \bibinfo {pages} {1766} (\bibinfo {year} {2012})}\BibitemShut
  {NoStop}%
\bibitem [{\citenamefont {Zhang}\ \emph {et~al.}(2014)\citenamefont {Zhang},
  \citenamefont {Feist}, \citenamefont {Rubio}, \citenamefont
  {Garc\'{i}a-Gonz\'{a}lez},\ and\ \citenamefont
  {Garc\'{i}a-Vidal}}]{ZhangRubio2014}%
  \BibitemOpen
  \bibfield  {author} {\bibinfo {author} {\bibfnamefont {P.}~\bibnamefont
  {Zhang}}, \bibinfo {author} {\bibfnamefont {J.}~\bibnamefont {Feist}},
  \bibinfo {author} {\bibfnamefont {A.}~\bibnamefont {Rubio}}, \bibinfo
  {author} {\bibfnamefont {P.}~\bibnamefont {Garc\'{i}a-Gonz\'{a}lez}}, \ and\
  \bibinfo {author} {\bibfnamefont {F.~J.}\ \bibnamefont {Garc\'{i}a-Vidal}},\
  }\href
  {http://www.scopus.com/inward/record.url?eid=2-s2.0-84908251496&partnerID=40&md5=02031155082296a7a3202593c41c2a76}
  {\bibfield  {journal} {\bibinfo  {journal} {Phys. Rev. B}\ }\textbf {\bibinfo
  {volume} {90}},\ \bibinfo {pages} {161407(R)} (\bibinfo {year}
  {2014})}\BibitemShut {NoStop}%
\bibitem [{\citenamefont {Zheng}\ \emph {et~al.}(2015)\citenamefont {Zheng},
  \citenamefont {Zhao}, \citenamefont {Manjavacas}, \citenamefont {McClain},
  \citenamefont {Nordlander},\ and\ \citenamefont {Halas}}]{Nordlander2015}%
  \BibitemOpen
  \bibfield  {author} {\bibinfo {author} {\bibfnamefont {B.~Y.}\ \bibnamefont
  {Zheng}}, \bibinfo {author} {\bibfnamefont {H.}~\bibnamefont {Zhao}},
  \bibinfo {author} {\bibfnamefont {A.}~\bibnamefont {Manjavacas}}, \bibinfo
  {author} {\bibfnamefont {M.}~\bibnamefont {McClain}}, \bibinfo {author}
  {\bibfnamefont {P.}~\bibnamefont {Nordlander}}, \ and\ \bibinfo {author}
  {\bibfnamefont {N.~J.}\ \bibnamefont {Halas}},\ }\href@noop {} {\bibfield
  {journal} {\bibinfo  {journal} {Nature Comm.}\ }\textbf {\bibinfo {volume}
  {6}},\ \bibinfo {pages} {7797} (\bibinfo {year} {2015})}\BibitemShut
  {NoStop}%
\bibitem [{\citenamefont {Guidez}\ and\ \citenamefont
  {Aikens}(2013)}]{Guidez2013}%
  \BibitemOpen
  \bibfield  {author} {\bibinfo {author} {\bibfnamefont {E.~B.}\ \bibnamefont
  {Guidez}}\ and\ \bibinfo {author} {\bibfnamefont {C.~M.}\ \bibnamefont
  {Aikens}},\ }\href {\doibase 10.1021/jp4059033} {\bibfield  {journal}
  {\bibinfo  {journal} {J. Phys. Chem. C}\ }\textbf {\bibinfo {volume} {117}},\
  \bibinfo {pages} {21466} (\bibinfo {year} {2013})}\BibitemShut {NoStop}%
\bibitem [{\citenamefont {Townsend}\ and\ \citenamefont
  {Bryant}(2014)}]{Townsend2014}%
  \BibitemOpen
  \bibfield  {author} {\bibinfo {author} {\bibfnamefont {E.}~\bibnamefont
  {Townsend}}\ and\ \bibinfo {author} {\bibfnamefont {G.~W.}\ \bibnamefont
  {Bryant}},\ }\href
  {http://www.scopus.com/inward/record.url?eid=2-s2.0-84908587158&partnerID=40&md5=1374822867712d66397dbeb7183c0bdc}
  {\bibfield  {journal} {\bibinfo  {journal} {J. Opt.}\ }\textbf {\bibinfo
  {volume} {16}},\ \bibinfo {pages} {114022} (\bibinfo {year}
  {2014})}\BibitemShut {NoStop}%
\bibitem [{\citenamefont {Guidez}\ and\ \citenamefont
  {Aikens}(2014)}]{Guidez2014Nanoscale}%
  \BibitemOpen
  \bibfield  {author} {\bibinfo {author} {\bibfnamefont {E.~B.}\ \bibnamefont
  {Guidez}}\ and\ \bibinfo {author} {\bibfnamefont {C.~M.}\ \bibnamefont
  {Aikens}},\ }\href {\doibase 10.1039/c4nr02225d} {\bibfield  {journal}
  {\bibinfo  {journal} {Nanoscale}\ }\textbf {\bibinfo {volume} {6}},\ \bibinfo
  {pages} {11512} (\bibinfo {year} {2014})}\BibitemShut {NoStop}%
\bibitem [{\citenamefont {Bursi}\ \emph {et~al.}(2014)\citenamefont {Bursi},
  \citenamefont {Calzolari}, \citenamefont {Corni},\ and\ \citenamefont
  {Molinari}}]{Bursi2014}%
  \BibitemOpen
  \bibfield  {author} {\bibinfo {author} {\bibfnamefont {L.}~\bibnamefont
  {Bursi}}, \bibinfo {author} {\bibfnamefont {A.}~\bibnamefont {Calzolari}},
  \bibinfo {author} {\bibfnamefont {S.}~\bibnamefont {Corni}}, \ and\ \bibinfo
  {author} {\bibfnamefont {E.}~\bibnamefont {Molinari}},\ }\href {\doibase
  10.1021/ph500269q} {\bibfield  {journal} {\bibinfo  {journal} {ACS
  Photonics}\ }\textbf {\bibinfo {volume} {1}},\ \bibinfo {pages} {1049}
  (\bibinfo {year} {2014})}\BibitemShut {NoStop}%
\bibitem [{\citenamefont {Krauter}\ \emph {et~al.}(2014)\citenamefont
  {Krauter}, \citenamefont {Schirmer}, \citenamefont {Jacob}, \citenamefont
  {Pernpointner},\ and\ \citenamefont {Dreuw}}]{Krauter2014}%
  \BibitemOpen
  \bibfield  {author} {\bibinfo {author} {\bibfnamefont {C.~M.}\ \bibnamefont
  {Krauter}}, \bibinfo {author} {\bibfnamefont {J.}~\bibnamefont {Schirmer}},
  \bibinfo {author} {\bibfnamefont {C.~R.}\ \bibnamefont {Jacob}}, \bibinfo
  {author} {\bibfnamefont {M.}~\bibnamefont {Pernpointner}}, \ and\ \bibinfo
  {author} {\bibfnamefont {A.}~\bibnamefont {Dreuw}},\ }\href {\doibase
  http://dx.doi.org/10.1063/1.4894266} {\bibfield  {journal} {\bibinfo
  {journal} {J. Chem. Phys.}\ }\textbf {\bibinfo {volume} {141}},\ \bibinfo
  {eid} {104101} (\bibinfo {year} {2014})}\BibitemShut {NoStop}%
\bibitem [{\citenamefont {Townsend}\ \emph {et~al.}(2015)\citenamefont
  {Townsend}, \citenamefont {Debrecht},\ and\ \citenamefont
  {Bryant}}]{Townsend2015}%
  \BibitemOpen
  \bibfield  {author} {\bibinfo {author} {\bibfnamefont {E.}~\bibnamefont
  {Townsend}}, \bibinfo {author} {\bibfnamefont {A.}~\bibnamefont {Debrecht}},
  \ and\ \bibinfo {author} {\bibfnamefont {G.~W.}\ \bibnamefont {Bryant}},\
  }\href {\doibase 10.1557/jmr.2015.232} {\bibfield  {journal} {\bibinfo
  {journal} {J. Mater. Res.}\ }\textbf {\bibinfo {volume} {30}},\ \bibinfo
  {pages} {2389} (\bibinfo {year} {2015})}\BibitemShut {NoStop}%
\bibitem [{Fet()}]{FetterWalecka}%
  \BibitemOpen
  \href@noop {} {}\bibinfo {note} {A. L. Fetter and J. D. Walecka, {\em Quantum
  Theory of Many-Particle Systems} (Dover Publications: Mineola, NY,
  2003).}\BibitemShut {Stop}%
\bibitem [{Mar()}]{MarquesGross}%
  \BibitemOpen
  \href@noop {} {}\bibinfo {note} {M. A. L. Marques, N. T. Maitra, F. M. S.
  Nogueira, E. K. U. Gross, and A. Rubio, (Eds.) {\em Fundamentals of
  Time-Dependent Density Functional Theory}, Springer series: Lecture Notes in
  Physics, Vol. 837 (Springer, Berlin, 2012).}\BibitemShut {Stop}%
\bibitem [{\citenamefont {Egri}(1985)}]{Egri1985}%
  \BibitemOpen
  \bibfield  {author} {\bibinfo {author} {\bibfnamefont {I.}~\bibnamefont
  {Egri}},\ }\href {\doibase http://dx.doi.org/10.1016/0370-1573(85)90085-7}
  {\bibfield  {journal} {\bibinfo  {journal} {Phys. Rep.}\ }\textbf {\bibinfo
  {volume} {119}},\ \bibinfo {pages} {363 } (\bibinfo {year}
  {1985})}\BibitemShut {NoStop}%
\bibitem [{SM()}]{SM}%
  \BibitemOpen
  \href@noop {} {}\bibinfo {note} {See Supplemental Material at [URL will be
  inserted by publisher].}\BibitemShut {Stop}%
\bibitem [{\citenamefont {Yan}\ \emph {et~al.}(2007)\citenamefont {Yan},
  \citenamefont {Yuan},\ and\ \citenamefont {Gao}}]{Yan2007}%
  \BibitemOpen
  \bibfield  {author} {\bibinfo {author} {\bibfnamefont {J.}~\bibnamefont
  {Yan}}, \bibinfo {author} {\bibfnamefont {Z.}~\bibnamefont {Yuan}}, \ and\
  \bibinfo {author} {\bibfnamefont {S.}~\bibnamefont {Gao}},\ }\href
  {http://www.scopus.com/inward/record.url?eid=2-s2.0-34547322086&partnerID=40&md5=a75581b50c06a72d9a99bc82fe854a36}
  {\bibfield  {journal} {\bibinfo  {journal} {Phys. Rev. Lett.}\ }\textbf
  {\bibinfo {volume} {98}},\ \bibinfo {pages} {216602} (\bibinfo {year}
  {2007})}\BibitemShut {NoStop}%
\bibitem [{\citenamefont {Chen}\ \emph {et~al.}(2013)\citenamefont {Chen},
  \citenamefont {Dyer}, \citenamefont {Li},\ and\ \citenamefont
  {Dixon}}]{Chen2013}%
  \BibitemOpen
  \bibfield  {author} {\bibinfo {author} {\bibfnamefont {M.}~\bibnamefont
  {Chen}}, \bibinfo {author} {\bibfnamefont {J.~E.}\ \bibnamefont {Dyer}},
  \bibinfo {author} {\bibfnamefont {K.}~\bibnamefont {Li}}, \ and\ \bibinfo
  {author} {\bibfnamefont {D.~A.}\ \bibnamefont {Dixon}},\ }\href
  {http://www.scopus.com/inward/record.url?eid=2-s2.0-84883574728&partnerID=40&md5=a011f519211f7a924fd1991a14c39770}
  {\bibfield  {journal} {\bibinfo  {journal} {J. Phys. Chem. A}\ }\textbf
  {\bibinfo {volume} {117}},\ \bibinfo {pages} {8298} (\bibinfo {year}
  {2013})}\BibitemShut {NoStop}%
\bibitem [{\citenamefont {Manjavacas}\ \emph {et~al.}(2013)\citenamefont
  {Manjavacas}, \citenamefont {Marchesin}, \citenamefont {Thongrattanasiri},
  \citenamefont {Koval}, \citenamefont {Nordlander}, \citenamefont
  {S\'{a}nchez-Portal},\ and\ \citenamefont {{Garc\'{\i}a de
  Abajo}}}]{Manjavacas2013}%
  \BibitemOpen
  \bibfield  {author} {\bibinfo {author} {\bibfnamefont {A.}~\bibnamefont
  {Manjavacas}}, \bibinfo {author} {\bibfnamefont {F.}~\bibnamefont
  {Marchesin}}, \bibinfo {author} {\bibfnamefont {S.}~\bibnamefont
  {Thongrattanasiri}}, \bibinfo {author} {\bibfnamefont {P.}~\bibnamefont
  {Koval}}, \bibinfo {author} {\bibfnamefont {P.}~\bibnamefont {Nordlander}},
  \bibinfo {author} {\bibfnamefont {D.}~\bibnamefont {S\'{a}nchez-Portal}}, \
  and\ \bibinfo {author} {\bibfnamefont {F.~J.}\ \bibnamefont {{Garc\'{\i}a de
  Abajo}}},\ }\href {\doibase 10.1021/nn4006297} {\bibfield  {journal}
  {\bibinfo  {journal} {ACS Nano}\ }\textbf {\bibinfo {volume} {7}},\ \bibinfo
  {pages} {3635} (\bibinfo {year} {2013})}\BibitemShut {NoStop}%
\bibitem [{\citenamefont {Lauchner}\ \emph {et~al.}(2015)\citenamefont
  {Lauchner}, \citenamefont {Schlather}, \citenamefont {Manjavacas},
  \citenamefont {Cui}, \citenamefont {McClain}, \citenamefont {Stec},
  \citenamefont {de~Abajo}, \citenamefont {Nordlander},\ and\ \citenamefont
  {Halas}}]{Lauchner2015}%
  \BibitemOpen
  \bibfield  {author} {\bibinfo {author} {\bibfnamefont {A.}~\bibnamefont
  {Lauchner}}, \bibinfo {author} {\bibfnamefont {A.~E.}\ \bibnamefont
  {Schlather}}, \bibinfo {author} {\bibfnamefont {A.}~\bibnamefont
  {Manjavacas}}, \bibinfo {author} {\bibfnamefont {Y.}~\bibnamefont {Cui}},
  \bibinfo {author} {\bibfnamefont {M.~J.}\ \bibnamefont {McClain}}, \bibinfo
  {author} {\bibfnamefont {G.~J.}\ \bibnamefont {Stec}}, \bibinfo {author}
  {\bibfnamefont {F.~J.~G.}\ \bibnamefont {de~Abajo}}, \bibinfo {author}
  {\bibfnamefont {P.}~\bibnamefont {Nordlander}}, \ and\ \bibinfo {author}
  {\bibfnamefont {N.~J.}\ \bibnamefont {Halas}},\ }\href {\doibase
  10.1021/acs.nanolett.5b02549} {\bibfield  {journal} {\bibinfo  {journal}
  {Nano Lett.}\ }\textbf {\bibinfo {volume} {15}},\ \bibinfo {pages} {6208}
  (\bibinfo {year} {2015})}\BibitemShut {NoStop}%
\bibitem [{\citenamefont {Zhao}\ \emph {et~al.}(2006)\citenamefont {Zhao},
  \citenamefont {Jensen},\ and\ \citenamefont {Schatz}}]{Zhao2006}%
  \BibitemOpen
  \bibfield  {author} {\bibinfo {author} {\bibfnamefont {L.}~\bibnamefont
  {Zhao}}, \bibinfo {author} {\bibfnamefont {L.}~\bibnamefont {Jensen}}, \ and\
  \bibinfo {author} {\bibfnamefont {G.}~\bibnamefont {Schatz}},\ }\href
  {\doibase 10.1021/ja0556326} {\bibfield  {journal} {\bibinfo  {journal} {J.
  Am. Chem. Soc.}\ }\textbf {\bibinfo {volume} {128}},\ \bibinfo {pages} {2911}
  (\bibinfo {year} {2006})}\BibitemShut {NoStop}%
\bibitem [{\citenamefont {Giannozzi}\ \emph {et~al.}(2009)\citenamefont
  {Giannozzi}, \citenamefont {Baroni}, \citenamefont {Bonini}, \citenamefont
  {Calandra}, \citenamefont {Car}, \citenamefont {Cavazzoni}, \citenamefont
  {Ceresoli}, \citenamefont {Chiarotti}, \citenamefont {Cococcioni},
  \citenamefont {Dabo}, \citenamefont {Dal~Corso}, \citenamefont
  {de~Gironcoli}, \citenamefont {Fabris}, \citenamefont {Fratesi},
  \citenamefont {Gebauer}, \citenamefont {Gerstmann}, \citenamefont
  {Gougoussis}, \citenamefont {Kokalj}, \citenamefont {Lazzeri}, \citenamefont
  {Martin-Samos}, \citenamefont {Marzari}, \citenamefont {Mauri}, \citenamefont
  {Mazzarello}, \citenamefont {Paolini}, \citenamefont {Pasquarello},
  \citenamefont {Paulatto}, \citenamefont {Sbraccia}, \citenamefont {Scandolo},
  \citenamefont {Sclauzero}, \citenamefont {Seitsonen}, \citenamefont
  {Smogunov}, \citenamefont {Umari},\ and\ \citenamefont
  {Wentzcovitch}}]{espresso}%
  \BibitemOpen
  \bibfield  {author} {\bibinfo {author} {\bibfnamefont {P.}~\bibnamefont
  {Giannozzi}}, \bibinfo {author} {\bibfnamefont {S.}~\bibnamefont {Baroni}},
  \bibinfo {author} {\bibfnamefont {N.}~\bibnamefont {Bonini}}, \bibinfo
  {author} {\bibfnamefont {M.}~\bibnamefont {Calandra}}, \bibinfo {author}
  {\bibfnamefont {R.}~\bibnamefont {Car}}, \bibinfo {author} {\bibfnamefont
  {C.}~\bibnamefont {Cavazzoni}}, \bibinfo {author} {\bibfnamefont
  {D.}~\bibnamefont {Ceresoli}}, \bibinfo {author} {\bibfnamefont {G.~L.}\
  \bibnamefont {Chiarotti}}, \bibinfo {author} {\bibfnamefont {M.}~\bibnamefont
  {Cococcioni}}, \bibinfo {author} {\bibfnamefont {I.}~\bibnamefont {Dabo}},
  \bibinfo {author} {\bibfnamefont {A.}~\bibnamefont {Dal~Corso}}, \bibinfo
  {author} {\bibfnamefont {S.}~\bibnamefont {de~Gironcoli}}, \bibinfo {author}
  {\bibfnamefont {S.}~\bibnamefont {Fabris}}, \bibinfo {author} {\bibfnamefont
  {G.}~\bibnamefont {Fratesi}}, \bibinfo {author} {\bibfnamefont
  {R.}~\bibnamefont {Gebauer}}, \bibinfo {author} {\bibfnamefont
  {U.}~\bibnamefont {Gerstmann}}, \bibinfo {author} {\bibfnamefont
  {C.}~\bibnamefont {Gougoussis}}, \bibinfo {author} {\bibfnamefont
  {A.}~\bibnamefont {Kokalj}}, \bibinfo {author} {\bibfnamefont
  {M.}~\bibnamefont {Lazzeri}}, \bibinfo {author} {\bibfnamefont
  {L.}~\bibnamefont {Martin-Samos}}, \bibinfo {author} {\bibfnamefont
  {N.}~\bibnamefont {Marzari}}, \bibinfo {author} {\bibfnamefont
  {F.}~\bibnamefont {Mauri}}, \bibinfo {author} {\bibfnamefont
  {R.}~\bibnamefont {Mazzarello}}, \bibinfo {author} {\bibfnamefont
  {S.}~\bibnamefont {Paolini}}, \bibinfo {author} {\bibfnamefont
  {A.}~\bibnamefont {Pasquarello}}, \bibinfo {author} {\bibfnamefont
  {L.}~\bibnamefont {Paulatto}}, \bibinfo {author} {\bibfnamefont
  {C.}~\bibnamefont {Sbraccia}}, \bibinfo {author} {\bibfnamefont
  {S.}~\bibnamefont {Scandolo}}, \bibinfo {author} {\bibfnamefont
  {G.}~\bibnamefont {Sclauzero}}, \bibinfo {author} {\bibfnamefont {A.~P.}\
  \bibnamefont {Seitsonen}}, \bibinfo {author} {\bibfnamefont {A.}~\bibnamefont
  {Smogunov}}, \bibinfo {author} {\bibfnamefont {P.}~\bibnamefont {Umari}}, \
  and\ \bibinfo {author} {\bibfnamefont {R.~M.}\ \bibnamefont {Wentzcovitch}},\
  }\href@noop {} {\bibfield  {journal} {\bibinfo  {journal} {J. Phys.: Condens.
  Matter}\ }\textbf {\bibinfo {volume} {21}},\ \bibinfo {pages} {395502}
  (\bibinfo {year} {2009})}\BibitemShut {NoStop}%
\bibitem [{\citenamefont {Perdew}\ \emph {et~al.}(1996)\citenamefont {Perdew},
  \citenamefont {Burke},\ and\ \citenamefont {Ernzerhof}}]{PBE1996}%
  \BibitemOpen
  \bibfield  {author} {\bibinfo {author} {\bibfnamefont {J.}~\bibnamefont
  {Perdew}}, \bibinfo {author} {\bibfnamefont {K.}~\bibnamefont {Burke}}, \
  and\ \bibinfo {author} {\bibfnamefont {M.}~\bibnamefont {Ernzerhof}},\ }\href
  {\doibase 10.1103/PhysRevLett.77.3865} {\bibfield  {journal} {\bibinfo
  {journal} {Phys. Rev. Lett.}\ }\textbf {\bibinfo {volume} {77}},\ \bibinfo
  {pages} {3865} (\bibinfo {year} {1996})}\BibitemShut {NoStop}%
\bibitem [{not()}]{note_cutoffs}%
  \BibitemOpen
  \href@noop {} {}\bibinfo {note} {The energy cutoff in the plane waves
  expansion of the single-particle wavefunctions and the charge density for the
  Na$ _{20} $ chain is 32 Ry and 384 Ry; for naphthalene, 25 Ry and 300 Ry; for
  the Ag$ _{20} $ cluster and the pyridine molecule, it is 33 Ry and 396 Ry,
  respectively.}\BibitemShut {Stop}%
\bibitem [{\citenamefont {Ge}\ \emph {et~al.}(2014)\citenamefont {Ge},
  \citenamefont {Binnie}, \citenamefont {Rocca}, \citenamefont {Gebauer},\ and\
  \citenamefont {Baroni}}]{Ge2014}%
  \BibitemOpen
  \bibfield  {author} {\bibinfo {author} {\bibfnamefont {X.}~\bibnamefont
  {Ge}}, \bibinfo {author} {\bibfnamefont {S.~J.}\ \bibnamefont {Binnie}},
  \bibinfo {author} {\bibfnamefont {D.}~\bibnamefont {Rocca}}, \bibinfo
  {author} {\bibfnamefont {R.}~\bibnamefont {Gebauer}}, \ and\ \bibinfo
  {author} {\bibfnamefont {S.}~\bibnamefont {Baroni}},\ }\href {\doibase
  http://dx.doi.org/10.1016/j.cpc.2014.03.005} {\bibfield  {journal} {\bibinfo
  {journal} {Comput. Phys. Commun.}\ }\textbf {\bibinfo {volume} {185}},\
  \bibinfo {pages} {2080 } (\bibinfo {year} {2014})}\BibitemShut {NoStop}%
\end{thebibliography}%

\end{document}